\documentstyle[12pt,a4]{article}

\begin{document}

\begin{center}

  {\Large\bf{
Charge separation in photosynthesis via a spin exchange coupling mechanism.
 }}

  \bigskip

  \bf{
    \underline{Sighart F. Fischer and P.O.J. Scherer}
     }

  Technische Universit\"at M\"unchen, Physikdepartment T38,
  D-85748 Garching, Germany

  E-mail: scherer@venus.t30.physik.tu-muenchen.de

  URL: http://jupiter.t30.physik.tu-muenchen.de/scherer/scherer.html

  \medskip

\end{center}

\bigskip

\section*{Abstract}
A new mechanism for the primary photoinduced charge separation in photosynthesis is proposed. It involves as real intermediate between the excited special pair state $P^*$ and the primary charge separated state $P^+ H_L^-$ a trip-trip-singlet $P^T B_L^T$,  which consists of a triplet on the dimer P and a further triplet on the monomer $B_L$. Both combine to a singlet. The electron transfer is caused by spin exchange couplings. The transient spectrum of the short lived intermediate, formerly taken as evidence for the charge transfer state $P^+ B_L^-$, is reinterpreted as a transient excitation of this trip-trip singlet.

\section{Introduction}
Photosynthetic systems convert photon energy into electrostatic energy within a few picoseconds with great efficiency. The initial charge separation takes place in a reaction center. Its basic structure (Deisenhofer et al 1984) (Fig. 1) seems to be preserved among all photosynthetic systems. Six pigments are arranged in an approximate $C_2$ symmetry forming an L and an M branch. Two make the special pair dimer. Two referred to as monomers are linked to it within van der Waals contact followed by two further pigments both being close to one of the monomers. The one on the L branch acts as the initial acceptor. In the special case of Rps. viridis, on which we want to base our calculations, the dimer P consists of two bacteriochlorophylls $P_L$ and $P_M$ noncovalently bound via $\pi$-orbital interactions of their pyrrol rings I. The monomers are two bacteriochlorophylls denoted $B_L$ and $B_M$ . They point towards the rings I of the dimer P with their rings III. The rings I of $B_L$ and $B_M$ face the corresponding rings I of the bacteriopheophytines $H_L$ and $H_M$ respectively. We will see that these basic structural features are essential to bring the energy location of the trip-trip singlet $P^T B_L^T$ below that of the initially excited dimer state $P^*$. Furthermore we show that the spin exchange couplings are particularly favorable for this arrangement. This is not so for the commonly postulated intermediate charge transfer state $P^+ B_L^-$ (Fischer and Scherer 1987; Scherer  and Fischer  1989a; Scherer  and Fischer  1989b; Holzapfel et al 1990; Scherer  1989). We predict it to lie above $P^*$ by about 0.5 e V. Our newly proposed intermediate $P^T B_L^T$ differs from the CT state $P^+ B_L^-$ by a compensating charge transfer excitation from $B_L$ to P so that two triplets are created which combine to a singlet. Coulomb and exchange energies overcompensate in this case the orbital excitation energy to bring the resulting trip-trip singlet $P^T B_L^T$ far below the CT state $P^+B_L^-$and even below $P^*$. The couplings between $P^T B_L^T$ to $P^*$ and to $P^+H_L^-$ involve two electron transfer overlaps, which differ from the Dexter exchange couplings relevant for excitation energy transfer only in the partitioning of different orbitals. The overall tunneling process from $P^*$ to $P^+H_L^-$ can be looked upon as an electron assisted tunneling in the sense that one electron from the HOMO of $B_L$ is moved to the LUMO of P and back once the transferred electron has passed $B_L$. This way the effective tunneling barrier is lowered (Fig.2).

	In order to check the relevance of this mechanism almost all experiments related to the charge separation, to recombination and to their magnetic and electric field dependence have to be reinterpreted. Particularly informative are changes in the kinetics due to mutations or chemical modifications of the pigments. In this paper we shall discuss briefly those cases, which have led to unexpected results within the standard model with $P^+B_L^-$ as intermediate. Of course the energetics of the trip-trip singlet $P^T B_L^T$ is sensitive to changes in a different way than  that of $P^+B_L^-$. Crudely speaking it is only  weakly dependent on changes of the electrostatic potentials but more sensitive to local distortions of the pigments structure. We will show that these characteristics of our model help to interpret the experimental findings including double mutants under a new perspective without multiparameter adjustments (Bixon et al 1996;).

\section{Energetics of charge transfer states}
In an attempt to interrelate structure and function for the reaction center it is instructive to analyze the energetics of the low lying charge transfer states with regard to the following five contributions:

\noindent
a) the local structure of the isolated pigments, which is used for the calculation of the ionization potentials and the electron affinities,

\noindent
b) the Coulomb interactions between the prosthetic groups, which are induced by a charge transfer transition,

\noindent
c) the polarization effects of these pigments which are due to charge reorganization under the influence of the induced Coulomb forces,

\noindent
d) the electrostatic polarization effects of those residues and water molecules which are in close neighborhood to the prosthetic groups,

\noindent
e) the long range electrostatic effects resulting from polar groups of the protein.

	In Fig. 1 those molecules are shown which are treated quantum mechanically. To get the local structural effects (a) the ionization potentials (IP) and the electron affinities (EA) were evaluated for the six isolated pigments $P_M$ , $P_L$ , $B_M$ , $B_L$ , $H_M$ and $H_L$. We used a semiempirical MO program of the INDO type with the parametrization similar to Zerner's ZINDO method (Thompson and Zerner 1990) with configuration interaction (Scherer  and Fischer 1990) including up to half a million states. The structure of Rps. viridis was taken from the protein data bank. The positions of the hydrogen atoms were optimized with the help of an MNDO program. The charge transfer (CT) induced Coulomb interactions (b) between a donor D and an acceptor A and the polar surrounding molecules M have three contributions

$$E_C(D^+,A^-,M)-E_C(D,A,M)=$$

\begin{equation} 
\sum_{
\begin{array}{c}
i \in D \\
j \in A \\
\end{array}
} {{\Delta q_i \Delta q_j} \over {r_{ij}}}
+\sum_{M \ne D,A} \sum_{
\begin{array}{c}
i \in M \\
j \in D \\
\end{array}
} {{q_i \Delta q_j} \over {r_{ij}}}
+\sum_{M \ne D,A} \sum_{
\begin{array}{c}
i \in M \\
j \in A \\
\end{array}
} {{q_i \Delta q_j} \over {r_{ij}}}
\end{equation}

$$=E_C(D^+,A^-) + E_C(D^+,M) + E_C(A^-,M)$$

While the first term stands for the common point monopole interactions of the CT induced charges between the donor D and acceptor A only, the other two account for the induced interactions of the donor and the acceptor with the other molecules M respectively. The energies for the set of CT states $P_L^+ P_M^-$, 
$P_L^+ B_L^-$, $P_M^+ B_L^-$ , $P_L^+ H_L^-$, $P_M^+ H_L^-$ , and those with L and M exchanged, are shown in Fig. 3. The monomer contribution, defined as IP-EA, is smallest for the CT state $P_L^+ H_L^-$, indicating that the positive charge is better localized on $P_L$ than on $P_M$. The calculated electron affinity of $H_L$ exceeds that of $B_L$ by 0.62 eV. This is much more than found in solution (0.3 eV, Fajer et al 1975) . To test our calculation we simulated the molecules in solution by relaxing their structure and attaching water to Mg instead of the histidine. This way we could approximately reproduce the experimental value of the difference in electron affinities and the measured absorption spectrum of $B_L^-$ in solution (Scherer 1990). So our calculations predict that the electron affinity of $B_L$ is much smaller in the reaction center than in solution. 

	The influence of the Coulomb energy $E_C$ (Eq. 1) introduces a sizable contribution to the asymmetry of the two symmetry related CT states $P_L^+H_L^-$ and $P_M^+ H_M^-$ with the first falling now below the latter by 0.44 eV. This is mainly due to the difference in the anion interaction of $H_L^-$ or $H_M^-$ with the respective ground state dipoles of $P_L$ and $P_M$ ($E_C(A^-,M)$ from (1)). The polarization $E_p$ was evaluated with a super molecule approach, first for the hexamer, consisting just of the pigments $P_L$, $P_M$, $B_L$, $B_M$ , $H_L$ and $H_M$ with the histidines attached to the Mg atoms. We then included the short range polarization effects $E_s$ resulting from the neighboring residues and water molecules. The largest short range effect evolved for the anion $H_L^-$ via GLU L104. The protein electrostatics $E_l$ has been evaluated by means of the DelPhi-program solving the Poisson Boltzmann equation (Scherer et al 1995). It lowers the CT state $P_L^+H_L^-$ so far that it falls below the $P^*$ state. This way $P^+H_L^-$ becomes energetically accessible for the photo-induced charge separation without activation. In fact it is the only CT state which fulfills this condition. The state $P_M^+ H_M^-$ is higher by 0.4 e V and the state $P_L^+ B_L^-$ is above $P^*$ by 0.7 eV (vertical energy difference). 

The internal CT states 
$P_L^+ P_M^-$ and $P_M^+ P_L^-$  are always below $P_L^+ B_L^-$ . This is easy to understand, since they experience a larger Coulomb attraction. For these states the Coulomb effect is shown only together with the hexamer polarization. The surroundings do not narrow the gap between $P^*$ and $P_L^+ B_L^-$ substantially. TYR M 208 reduces it  only by 0.11 e V (Scherer et al 1995; Alden et al 1996) and a similar reduction was found for the water molecules (Scherer et al 1995). Interestingly the energy of $P_L^+ B_L^-$ remained almost unaffected by the long range electrostatic interactions of the protein. This conclusion is in line with the results of Marchi et al (1993). We could not reproduce certain results based on dynamics simulations by Warshel et al$^1$ , see also (Parson et al 1990; Alden et al 1995). Apparently they allowed for drastic changes in the structure. Following our theoretical prediction for the energy of $P^+ B_L^-$, we like to rule out this state as a real intermediate. Moreover, the electron transfer coupling between $P^*$ and $P^+ B_L^-$ is so small  that a superexchange mechanism cannot be operative either. On the other hand, we will see that the trip-trip singlet $P_L^T B_L^T$  can fulfill the basic requirements for the energetics and the couplings. 

\section{The trip-trip singlet $P^T B_L^T$}

Double excited singlet states resulting from two locally excited triplets play a major role in biological photosystems.  They are optically forbidden but can be populated via rapid internal conversion processes. In polyenes, they are assigned as $A_g$ states. For larger chains their energy falls below that of the lowest optically active $B_{1u}$ state (Hudson and Kohler 1972; Hudson and Kohler 1973). In biological systems the carotenoids make use of the rapid conversion process. A similar $A_g$ state plays an important role in the primary isomerization of the retinal chromophore in bacteriorhodopsin (Schulten et al 1995; Takeuchi and Tahara 1997). Another application is found in the isomerization of provitamin D (Sobolewski 1994; Fuß et al 1997).

	We were recently able to assign a trip-trip singlet within the reaction center, there denoted as double triplet $P_L^TP_M^T$ (Fischer and Scherer 1997). It has been observed in the transient excitation spectrum of $P^*$ (Wynne et al 1996). It gets intensity mostly from the internal CT state $P_L^+  P_M^-$ . Its energy location is 0.35 e V above $P^*$. 

	Instead of presenting already here detailed quantum calculations on the corresponding trip-trip singlet $P^T B_L^T$, we like to relate its energy to the experimentally detected trip-trip singlet $P_L^TP_M^T$ in a semiempirical way. Within the dimer the triplet energy is lower for $P_L$ compared to $P_M$. This follows from the analysis of ADMR studies (Hoff and Vrieze 1996). The triplets of $P_M$ and of $B_L$ are according to our calculations similar in energy, so that the sums of the localized triplet energies for $P_L^T P_M^T$ and for $P_L^T B_L^T$ respectively, should be about the same. Their induced Coulomb energies, which result from changes in the charge distribution due to the trip-trip-singlet excitations, are very different. For $P_M^T P_L^T$ we found a strong repulsion of 0.57 eV due to the charge shifts into the rings I from the rings III and due to the head to head arrangement of the dimer. For $P_L^T B_L^T$ we have a head-tail arrangement which results in a weak  Coulomb attraction of -0.03 e V. The state $P_M^T B_L^T$ induces an even larger Coulomb attraction of - 0.48 eV. This energy gain is overcompensated by the lower local triplet energy of $P_L$ for the $P_L^T B_L^T$ state. Since for $P^T B_L^T$ the triplet component of P should be largely localized on $P_L$ as it is for $P^T$ alone (Hoff and Vrieze 1996) we predict the final energy of $P^T B_L^T$, which incorporates CI interaction, close to $P^*$.

	In Fig.4 the calculated transient spectrum of the trip-trip singlet $P^T B_L^T$ is shown together with the ground state spectrum and experimental results (Dressler et al 1990). They show the change in absorbance for the intermediate between $P^*$ and $P^+H_L^-$ relative to the ground state absorption. The experimental spectrum applies to the two step transfer scheme, whereby the second step is the faster. The calculation is based on the tetramer $B_M$ $P_L$ $P_M$ $B_L$ with the histidines attached to the Mg-atoms. Excitations on $B_M$ are not incorporated. They should largely cancel out in the difference spectrum. In the ground state spectrum the strong dimer excitation $P^*$ is predicted at 990 nm in quite good agreement with experiments. The next band polarized almost perpendicular (83$^o$) to $P^*$ (0$^o$) is the so called upper dimer band. It is somewhat too high in energy in comparison with experiments and so is the $B_L^*$ excitation, which has the polarization angle of 33$^o$. The proper shifts to lower energies are indicated by horizontal arrows in Fig. 4.

	The transient $P^T B_L^T$ absorption shows a transition at 1060 nm, which correlates nicely with the observed transition at 1050 nm (Dressler et al 1990; Zinth et al 1996). Also its polarization of 30$^o$ is in line with the observed 28$^o$. In our calculation it is composed of two accidentally degenerate transitions on the dimer. Both are difficult to assign orbital wise, since many CI components contribute. The situation is similar to the trip doublet of $P^+$ (Fischer and Scherer 1997) where the charge is replaced by the triplet excitation on $P_L$. The next transition at 915 nm is localized on $B_L$. It should not be as broad as those on P, which have more charge transfer character. Therefore it should be worth while to search for such a state in this energy regime to test our model. The dominating strong transition at 750 nm is largely the
 $P_M^*$ excitation in the presence of the triplet on $P_L$. Together with the transition of $B_L$ the main change in the sign of the observed spectrum around 800 nm can be explained once we allow for the shift of $B_L^*$ and the upper dimer band, which are both experimentally justified. In addition a shift of the $B_M$ * excitation (not in the calculation) to higher energies is expected which might contribute to the amplitudes of the two components around 820 nm. Finally there are the two almost perpendicular polarized transitions at 680 nm followed by one localized on $B_L$ at 665 nm with a small polarization angle of 11$^o$. This is also consistent with the observation in this frequency regime. The transitions at 1050 nm and at 650 nm have been taken as evidence for the occurrence of $B_L^-$.(Zinth et al 1996). We argue this coincidence might be in some parts accidental, since the spectrum of $B_L^-$ in solution is not necessarily representative for the spectrum of the $P^+ B_L^-$ state in Rps. viridis. To prove this point, we evaluated the transient absorption spectrum of $P^+ B_L^-$ (Fig 5a) for the isolated pigments P and $B_L$ in the presence of the counter charge. There is no transition localized on $B_L$ around 1050 nm. Apparently the spectrum is shifted relative to the solution spectrum (Fajer et al 1973) to higher energies. For a relaxed structure with a water molecule attached we do find the low energy transition consistent with earlier calculations (Scherer 1990) in close agreement with the solution experiments (Fajer et al 1973) (Fig. 5b). We think a shift by 0.6 eV is significant and it is outside the uncertainties of the calculation and the structure.

	For the coupling between $P^*$ and $P^TB_L^T$  we have only preliminary results. $P^T B_L^T$ can be approximated by the localized state $P_L^T B_L^T$ . It couples to the excitonic component of $P^*$ which is localized on $P_L$ and to the internal charge transfer states of the dimer. The dominant matrix element of the electron-electron interaction reads

\begin{equation}
V(P_L^TB_L^T,P_L^*)=\sqrt{{2} \over {3}} (V_{P*B,B*P*} -V_{PB,B*P})
\end{equation}
 
We evaluated the relevant two center integrals with atomic Hartree-Fock wave functions and obtained for V as one major contribution 5 cm-1, not so different from the one particle coupling for the $P* \rightarrow P^+ B_L^-$ based on the INDO approximation (10 cm-1, Fischer  and Scherer  1987; Scherer  and Fischer  1989a; Scherer  and Fischer  1989b). In (2) there are almost no interferences between the dominating atomic contributions. This makes the coupling competitive to the one particle coupling of $P^*$ to $P^+ B_L^-$, which contains strong interferences. The $P^T B_L^T \rightarrow P^+ H_L^-$ coupling involves three center integrals which we cannot handle at this time sufficiently well. It must be efficient in order to assure the fast kinetics for the second step. The second rate process becomes in our model the first charge separating step describing the simultaneous shift of two electrons from P to $B_L$ and from $B_L$ to $H_L$ respectively. This implies that modifications on P can affect this second rate process not only via changes in the oxidation potential of P, but also via a change in the coupling. A similar coupling might be responsible for the very rapid charge separation following a $B_L^*$ excitation, which can bypass $P^*$ as predicted by us (Fischer and Scherer 1987) and recently verified experimentally (Van Brederode et al 1997) for a mutant.

\section{Charge separation for mutants and modified reaction centers}

As mentioned above the time dependent transient spectra for native reaction centers show two clearly separated kinetics for the formation of the charge separated state $P^+H_L^-$ . In the spirit of the two step model (Zinth et al 1996) the rate determining step $P* \rightarrow P^+ B_L^-$ is now to be replaced by the $P* \rightarrow  P^T B_L^T$ transition  followed by the faster step for the $P^+H_L^-$ formation. To test the model it is informative to analyze the energetics of modified reaction centers and to test implications for the kinetics. Modifications or mutations can change the energetics of the charge transfer states $P^+ B_L^-$ and $P^+H_L^-$ in three ways - via changes of the oxidation potential of $P^+$ or via changes of the redox potentials of $P^+ B_L^-$ or those of $P^+H_L^-$. The corresponding changes of the state $P^T B_L^T$ should be much smaller in all these cases, since energy differences between the LUMO's and HOMO's react less sensitively to such modifications than the energies themselves. 

	Within the first group two double mutants are of particular interest. For L 151 L$\rightarrow$ H  +  M 160 L$\rightarrow$ H (Woodbury et al 1994) two additional hydrogen bonds are introduced for the dimer, which raises the oxidation potential by 0.14 eV. The double mutant M 202 H$\rightarrow$L  +  L 131 L
$\rightarrow$ H (Laporte et al 1996) introduces a bacteriopheophytin in place of $P_M$ and adds a hydrogen bond on the keto group of $P_L$. It raises the oxidation potential by 0.26 eV. For these systems the charge separation process is still operative and an intermediate seems to evolve on the 4 ps time scale better detected for the first double mutant, which has signatures similar to the transient of $P^*$ (Woodbury et al 1994). The final charge separation is weakly activated. The common model needs to invoke for these systems a super-exchange matrix element for the coupling, which should contain a reduction factor of three orders of magnitude in this rate process compared to the two step process. There would be no explanation for an intermediate. This is not compatible with the observation. Within our model we could allow for a small up shift of the energy of the $P^T B_L^T$ state and keep it still as a real intermediate. Its transient spectrum can become consistent with the observed spectral changes once the rapid bypass process (Fischer and Scherer 1987; Scherer  and Fischer  1989b; Van Brederode 1997) is incorporated for a $B_L^*$ excitation. 

	Reduction of the oxidation potential by 0.08 eV has been achieved for the L168 H$\rightarrow$F mutant for Rps viridis (Arlt et al 1996). This change led to a threefold increase of the rate determining step, which enhances the population of the intermediate. To us it seems difficult to envision such an increase in the $P^*\rightarrow P^+ B_L^-$ transition since that coupling is localized in the region of closest approach between P and $B_L$. Our new coupling mechanism becomes sensitive however to vibronically induced couplings of the acetyl group of $P_L$, since this modifies the internal CT character of $P^*$.

	The variations in the kinetics caused by strong modifications of the redox potential of $P^+ B_L^-$ are also difficult to envision within the common model. The replacement of $B_L$ by a bacteriopheophytin (Zinth  et al 1996) should lower the redox potential by at least 0.3 eV (Fajer et al 1975). We predict an even larger effect due to structural relaxation. This value is larger than the full energy difference between $P^*$ and $P^+H_L^-$, so that $P^+ B_L^-$ (BPheo) should form a trap. The same should apply for the Ni replacement (H\"aberle et al 1996) of Mg on $B_L$. A change of the redox potential of 0.29 eV is expected on the basis of  measurements in solution. Within our model in both cases only minor changes of the energy of the $P^T B_L^T$ state are predicted which is consistent with the observation of an almost identical decay kinetics for $P^*$ (H\"aberle et al 1996). For the recombination the state $P^+ B_L^-$ (BPheo) might play in our model the same role of an equilibrated state as postulated in the literature (Zinth  et al 1996), since we would predict its energy location close to that of $P^+H_L^-$. 

	Heterodimers are mutations which can affect the energetics of the $P^T B_L^T$ state in a well defined way (McDowell et al 1991). Since the triplet energy of bacteriopheophytin is higher than that of bacteriochlorophyll and since the triplet of $P^T$ is localized on $P_L$ ,we predict a stronger increase of the energy $P^T B_L^T$ for the L heterodimer as compared to the M heterodimer. This is also consistent with the observations (McDowell et al 1991). The change of the redox potential of $P^+H_L^-$ by the replacement of $H_L$  by a pheophytin for Rb. sphaeroides (Schmidt et al 1995) does not require a new interpretation as long as the $P^+ B_L^-$ state is replaced by our $P^T B_L^T$ state. In analogy we would conclude that the trip-trip-singlet should be about 450 $cm^{-1}$ below $P^*$.

	For many other mutations such as the 3 vinyl - 13 OH bacteriochlorophyll substitution of $B_L$ (Finkele et al 1992; Nagarajan et al 1990; Finkele et al 1990; Shochat et al 1994) the variation of the energetics of $P^+ B_L^-$ and $P^T B_L^T$ should change in the same direction even though different couplings are responsible.

\section{Summary}

Our newly introduced intermediate trip-trip singlet $P^T B_L^T$ may bring us to a better understanding of the very special features of the reaction center. Within this model, we argue that the real trick for the photo-induced charge separation accomplished in the evolution process by the construction of the reaction center is the rapid delocalization of the excitation energy of $P^*$ over the dimer and the monomer in form of the trip-trip singlet $P^T B_L^T$. The formation of the charge separated state $P^+H_L^-$ avoids the appearance of a radical pair between neighboring molecules. This way the recombination is suppressed and the Coulomb attraction between the newly born radical pair is already strongly shielded by the intermediate $B_L$. Moreover, this mechanism avoids large nuclear reorganization for the initial step, since the state $P^T B_L^T$ undergoes no strong dipole change with respect to $P^*$. For $P^+H_L^-$ the Coulomb interaction is sufficiently shielded. We have shown that the detailed engineering makes largely use of the electrostatics, most important of that between the prosthetic groups. The unidirectionality is partly caused by the surrounding but mostly by the asymmetry within the dimer which localizes the triplet and to some extent also the positive charge on $P_L$. 

	An additional way to test the model might come from the magnetic field effects. The couplings for the triplet recombination are altered in this model and different predictions result for the Ni mutant (H\"aberle et al 1996) within the two models. In particular the superexchange coupling for recombination of the radical pair $P^+H_L^-$ to the triplet $P^T$ should be differently affected.

\section*{Acknowledgment} 
This work has been supported by the Deutsche Forschungsgemeinschaft (SFB 143 and SFB 533).

\section*{Footnotes}

1 A.Warshel kindly provided us with a protein conformation from his dynamics simulation which put the energy of $P^+ B_L^-$ below that of $P^*$. In this structure some residues were displaced by several \AA \quad and the Mg-HIS bond at $P_L$ was interrupted by an H-atom.

\section*{References}

{\footnotesize

\begin{description}

\item 
Alden R G, Parson W W, Chu Z T, Warshel A (1995) Calculations of electrostatic energies in photosynthetic reaction centers. J.Am.Chem.Soc. 117: 12284-12298

\item
Alden R G, Parson W W, Chu Z T, Warshel A. (1996) Orientation of the OH dipole of Tyrosine (M)210 and its effect on electrostatic energies in bacterial reaction centers. J.Phys.Chem. 100: 16761-16770

\item 
Arlt T, Bibikova M, Penzkofer H, Oesterhelt D, Zinth W (1996) Strong acceleration of primary photosynthetic electron-transfer in a mutated reaction center of Rhodopsendomanus-viridis. J.Phys.Chem. 100: 12060-12065

\item
Bixon M, Jortner J, Michel-Beyerle M E (1996) Energetics of the primary charge separation in bacterial photosynthesis. In: Michel-Beyerle M E (ed) The Reaction Center of Photosynthetic Bacteria, Springer, Berlin, pp 287-296

\item
Deisenhofer J, Epp O, Miki K, Huber R, Michel H (1984) X-ray structure analysis of a membrane protein complex: Electron density map at 3 A resolution and a model of the chromophores of the photosynthetic reaction center from Rhodopseudomonas viridis. J.Mol.Biol. 180: 385-398 

\item
Dressler K, Finkele U, Lauterwasser C, Hamm P, Holzapfel W, Buchanan S, Kaiser W, Michel H, Oesterhelt D, Scheer H, Stilz H U, Zinth W (1990) Similarities of the primary charge separation process in the photosynthesis of Rhodobacter sphaeroides amd Rodopseudomanas viridis. In: Michel-Beyerle M E (ed) Reaction Center of Photosynthetic Bacteria, Springer, Berlin, pp 135-140

\item
Fajer J, Brune D C, Davis M S, Forman A, Spaulding L D (1975) Primary charge separation in bacterial photosynthesis: Oxidized chlorophylls and reduced pheophytin. Proc.Nat.Acad.Sci. USA 72: 4956-4960

\item
Fajer J, Borg D C, Forman A, Dolphin D, Felton R H (1973) Anion radical of bacteriochlorophyll. J.Am.Chem.Soc. 95: 2739-2741

\item
Finkele U, Lauterwasser C, Zinth W, Gray K A, Oesterhelt D (1990) Role of tyrosine M210 in the initial charge separation in reaction centers of Rhodobacter sphaeroides. Biochemistry 29: 8517-8521

\item
Finkele U, Lauterwasser C, Struck A, Scheer H, Zinth W (1992) Primary electron transfer kinetics in bacterial reaction centers with modified bacteriochlorophylls at the monomeric sites BA,B . Proc. Natl Acad. Sci 89: 9514-9518

\item
Fischer S F, Scherer P O J (1987) On the early charge separation and recombination processes in bacterial reaction centers. Chem Phys 115:151-158.

\item
Fuss W, Hering P., Kompa K L, Lochbrunner S, Schikarski T, Schmid W E, Trushin S A (1997) Ultra photochemical pericyclic reactions and isomerizations of small polyenes. Ber. Bunsenges. Phys. Chem. 101: 500-509

\item
H\"aberle T, Lossau H, Friese M, Hartwich G, Ogrodnik A, Scheer H, Michel-Beyerle M E (1996) Ultrafast electron and excitation energy transfer in modified photosynthetic reaction centers from Rhodobacter sphaeroides, In: Michel-Beyerle M E (ed) The Reaction Center of Photosynthetic Bacteria. Springer, Berlin, pp 239-254

\item
Holzapfel W, Finkele U, Kaiser W, Oesterhelt D, Scheer H, Stiltz H U, Zinth W (1990) Initial electron-transfer in the reaction center from Rhodobacter sphaeroides. Proc. Natl. Acad. Sci. U.S.A. 87: 5168-5172

\item
Hudson B, Kohler B E (1972) A low-lying weak transition in the polyene (, (-diphenyloctatetraene. Chem.Phys.Lett. 14: 299-304

\item
Hudson B, Kohler B E (1973) Polyene spectroscopy: The lowest energy excited singlet state of diphenyloctatetraene and other linear polymers. J.Chem.Phys. 59: 4984-5002

\item
Laporte L L, Palaniappan V, Davis D G, Kirmaier C, Schenck C C, Holten D and Bocian D F (1996) Influence of electronic asymmetry on the spectroscopic photodynamic properties of primary electron-donor pairs in the photosynthetic reaction center. J.Phys.Chem. 100: 17696-17707

\item
Marchi M, Gehlen J N, Chandler D, Newton M (1993) Diabatic surfaces and the pathway for primary electron-transfer in photosynthetic reaction centers. J.Am.Chem.Soc. 115: 4178-4190

\item
McDowell L M, Gaul D, Kirmaier C, Holten D, Schenck C C (1991) Investigation into the source of electron transfer asymmetry in bacterial reaction centers. Biochemistry 30: 8315-8322

\item
Nagarajan V, Davis D, Parson W W, Gaul D, Schenck C (1990) Effect of specific mutations of tirosine-(M)210 on the primary photosynthetic electron-transfer process in Rhodobacter sphaeroides. Proc.Natl.Acad.Sci. USA 87: 7888-7892

\item
Parson W W, Chu Z T, Warshel A (1990) Electrostatic control of charge separation in bacterial photosynthesis. Biochim.Biophys.Acta 1017: 251-272

\item
Scherer P O J, Fischer S F (1989a) Long-range electron transfer within the hexamer of the photosynthetic reaction center Rhodopseudomonas viridis. J Phys Chem 93: 1633-1637

\item
Scherer P O J, Fischer S F (1989b) Quantum treatment of the optical spectra and the initial electron transfer process within the reaction center of Rhodopseudomonas viridis. Chem Phys 131: 115-127

\item
Scherer P O J (1989) Theoretical models for electron transfer and time resolved spectroscopy of the bacterial reaction center. Bulletin de la Societe Royale des Sciences de Liege 58e annee 3-4 247:247-252

\item
Scherer P O J (1990) Multiple exited states of photosynthetic reaction centers. In: Michel-Beyerle M E (ed) Reaction Center of Photosynthetic Bacteria, Springer, Berlin, pp 401-408

\item
Scherer P O J, Fischer S F (1990) Electronic excitations and electron transfer coupling within the bacterial reaction center based on an INDOS/S-CI supermolecule approach including 615 atoms. In: Jortner J and Pullman P (eds) Perspectives in Photosynthesis, Kluwer Academic publishers, Amsterdam, pp 361-370

\item
Scherer P O J, Scharnagl C, Fischer S F (1995) Symmetry Breakage in the Electronic Structure of Bacterial Reaction Centers. Chem. Phys. 197: 333-341

\item
Scherer P O J, Fischer S F (1997) Interpretation of a low-lying excited state of the reaction center of Rb. sphaeroides as a double triplet. Chem. Phys. Lett. 268:133-142
(lanl-physics/9702011 )

\item
Schmidt S, Arlt T, Hamm P, Huber H, N\"agele T, Wachtveitl J, Meyer M, Scheer H, Zinth W (1995) Primary electron-transfer dynamics in modified bacterial reaction centers containing pheophytin-A instead of Bacteriopheophytin-A. Spectrochim. Acta 51A: 1565-1578

\item
Schulten K, Humphrey W, Logunov I, Sheves M, Xu D (1995) Molecular Dynamics Studies of Bacteriorhodopsin´s photocycles. Israel Journal of Chemistry 35: 447-464

\item
Shochat S, Arlt T, Francke C, Gast P, van Noort P L, Otte S C, Schelvis H P M, Schmidt S, Vijgenboom E, Vrieze J, Zinth W, Hoff A J (1994) Spectroscopic characterization of reaction centers of the (M)Y210W mutant of the photosynthetic bacterium Rhodobacter-sphaeroides. J.Photosynth.Res. 40: 55-66

\item
Sobolewski A L, Domcke W (1994) Theoretical investigation of potential energy surfaces relevant for excited-state hydrogen transfer in o-hydroxybenzaldehyde. Chem.Phys. 184: 115-124

\item
Takeuchi S, Tahara T (1997) Ultrafast fluorescence study of the excited singlet-state dynamics of all-trans-retinal. J.Phys.Chem. A 101: 3052-3060

\item
Thompson M A, Zerner M C (1990) Effect of a polarizable medium on the charge-transfer states of the photosynthetic reaction center from Rhodopseudomonas viridis. J.Am.Chem.Soc. 112: 7828-7830

\item
Van Brederode M E, Jones M R, Van Grondelle R (1997) Fluorescence excitation spectra of membrane-bound photosynthetic reaction center of Rhodobacter sphaeroides in which the tyrosine M210 residue is replaced by tryptophan: evidence for a new pathway of charge separation. Chem.Phys.Lett. 268: 143-149.

\item
Vrieze J, Hoff A J (1996) Interactions between chromophores in reaction centers of purple bacteria - A reinterpretation of the triplet-minus-singlet spectra of Rhodobacter sphaeroides R26 and Rhodopsendomonas-viridis. Biochim.Biophys.Acta 1276: 210-220

\item
Warshel A, Chu Z T, Parson W W (1995) On the energetics of the primary electron-transfer process in bacterial reaction centers. Photochem.Photobiol. A.Chem. 82: 123-128

\item
Woodbury N W, Peloquin J M, Alden R G, Lin X, Lin S, Taguchi A K W, Williams J C, Allen J P (1994) Relationship between thermodynamics and mechanism of photoinduced charge separation in reaction centers from Rhodobacter sphaeroides. Biochemistry 33: 8101-8112

\item
Wynne K, Haran G, Reid G D, Moser C C, Dutton P L Hochstrasser R M (1996) Femtosecond infrared-spectroscopy of low-lying excited states in reaction centers of Rhodobacter sphaeroides. J.Phys. Chem. 100: 5140-5148

\item
Zinth W, Arlt T, Wachtveitl J (1996) The primary processes of bacterial photosynthesis - ultrafast reactions for the optimum use of light energy. Ber. Bunsenges. Phys.Chem. 100: 1962-1966.

\item
Zinth W, Arlt T, Schmidt S, Penzkofer H, Wachtveitl J, Huber H, N\"agele T, Hamm P, Bibikova M, Oesterhelt D, Meyer M, Scheer H (1996) The first femtoseconds of primary photosynthesis - The processes of the initial electron transfer reaction. In: Michel-Beyerle M E (ed) The Reaction Center of Photosynthetic Bacteria. Springer, Berlin, pp 159-173

\end{description}
}

\section*{Figure Captions}

\begin{description}

\item
Figure 1:	The quantum chemically treated part of the reaction center Rps. viridis is shown. It consists of the four bacteriochlorophylls $B_M$ , $P_L$, $P_M$ and $B_L$, the two bacteriopheophytines $H_M$ and $H_L$, the quinon $Q_A$ and several protein residues which are in close contact.

\item
Figure 2:	The electron assisted electron transfer mechanism is visualized. The solid arrows refer to the initial presumably slower (3.5 ps) process 
$P^* \rightarrow P^T B_L^T$, which invokes a two electron exchange between P and $B_L$. The dashed arrows give the  second faster (0.65 ps) charge separating step as a simultaneous transfer of an electron  from $B_L$ to $H_L$ and one from $B_L$ to P. The orbital energies refer to the neutral ground state of the hexamer from Fig. 1.

\item
Figure 3:	The calculated energies of the charge transfer states are shown.  Several contributions are presented separately. Ionisation potentials of the donor IP and electron affinities of the acceptor EA are estimated from the calculated MO energies of the isolated chromophores. The Coulomb energy $E_C$ results from the corresponding electron densities of the MO's. The polarization  contribution $E_p$ is evaluated from a calculation including the six chromophores as a supermolecule. Short range interactions $E_s$ refer to the residues shown in Fig.1 and are treated explicitly whereas long range electrostatic effects $E_l$ are treated in a continuum approximation.

\item
Figure 4:	The calculated transient difference spectrum of $P^T B_L^T$ (bars) is compared with experimental values (circles) from (Dressler et al 1990). The numbers show the calculated polarization angles relative to the dimer band $P^*$. The horizontal arrows, indicate the energy shifts needed for the experimental assignment.

\item
Figure 5a:	Calculated transient spectra for the CT state $P^+B_L^-$ based on the structure of Rps. viridis. The angles shown for the transitions of $B_L^-$ and $P^+$ are relative to the dimer transition $P^*$. The spectra of $P^+$(thin bars and diamonds)  and $B_L^-$ (thick bars and triangles) are calculated in the presence of the corresponding counter charge. They are superimposed.
\item
Figure 5b:	The spectrum of $P^+$ (thin bars and diamonds) is superimposed with the spectrum calculated for a relaxed $BChl^-H_2O$ anion complex (thick bars and triangles).

\end{description}

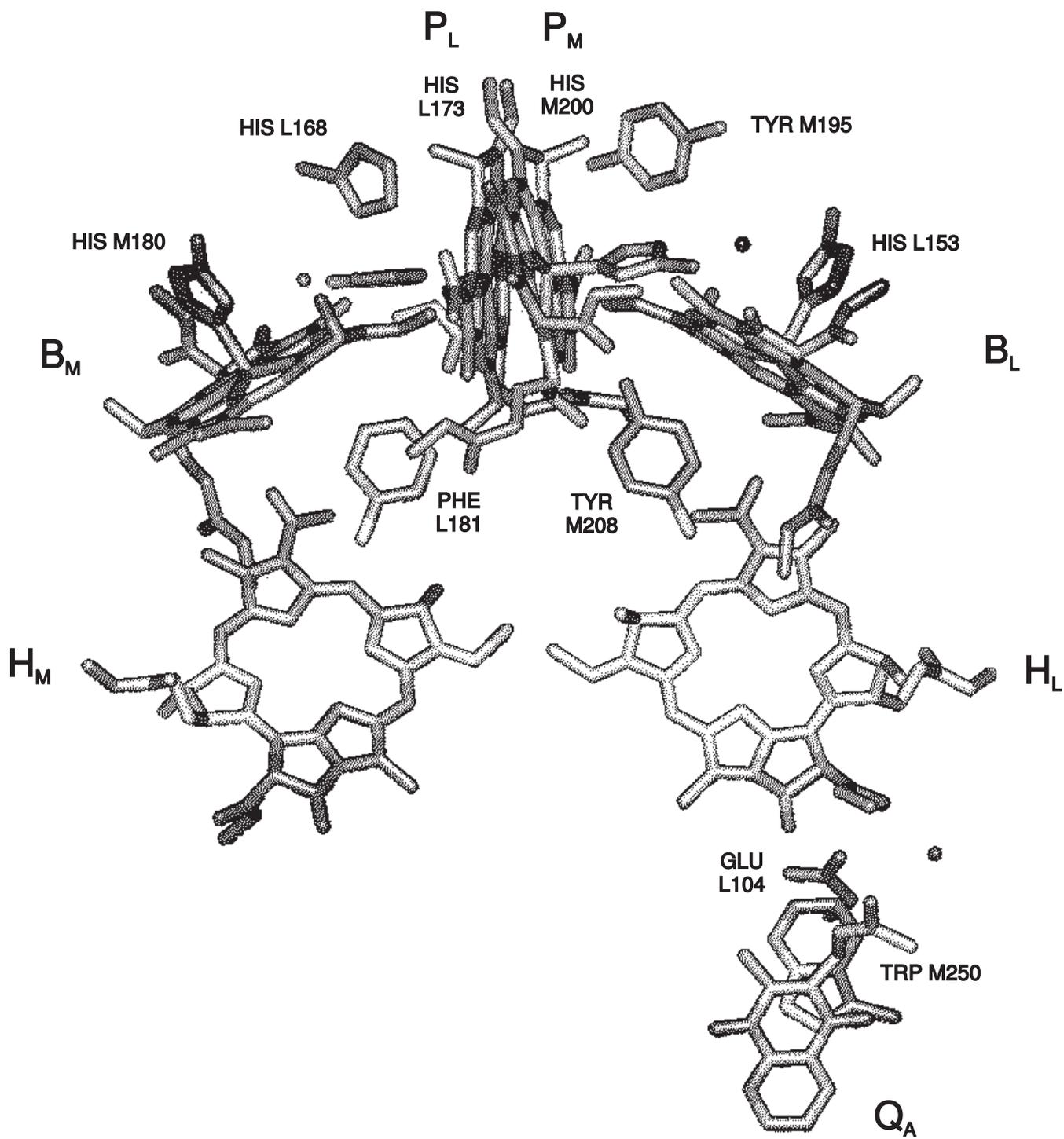
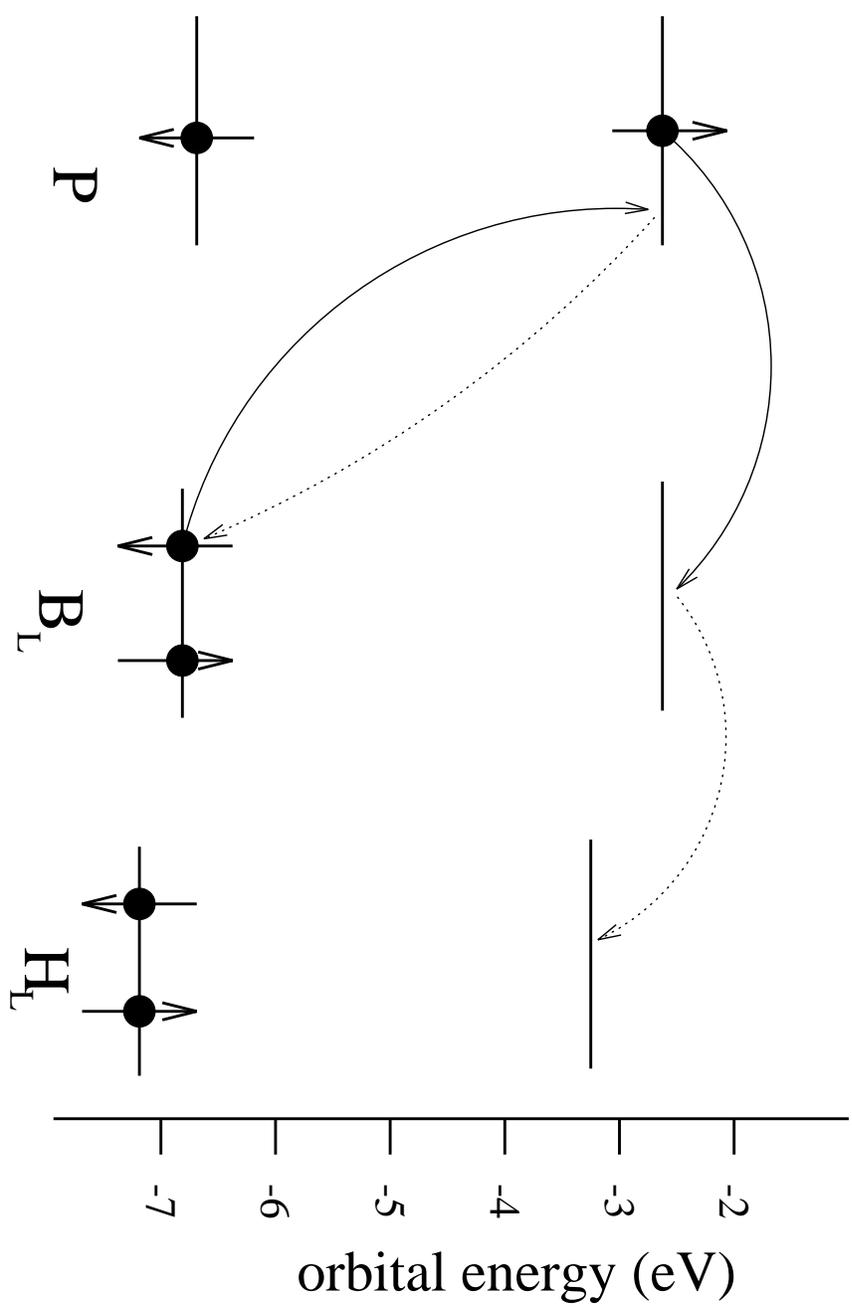
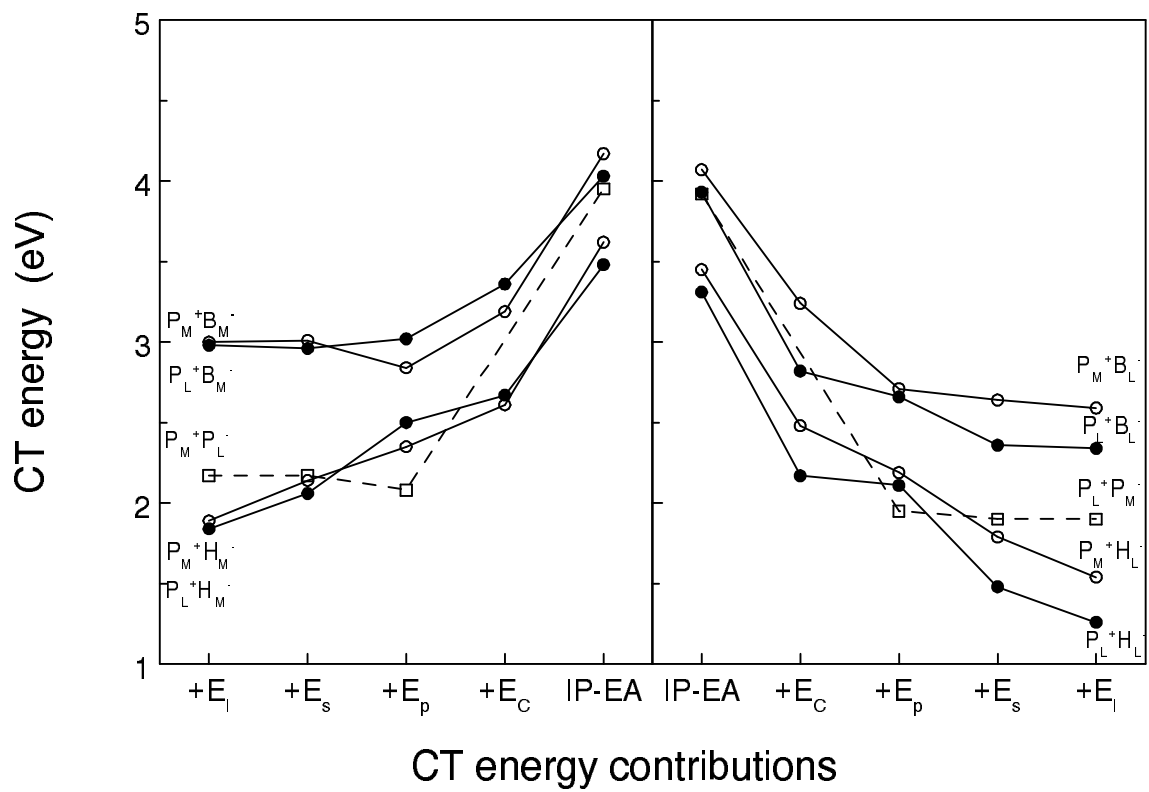
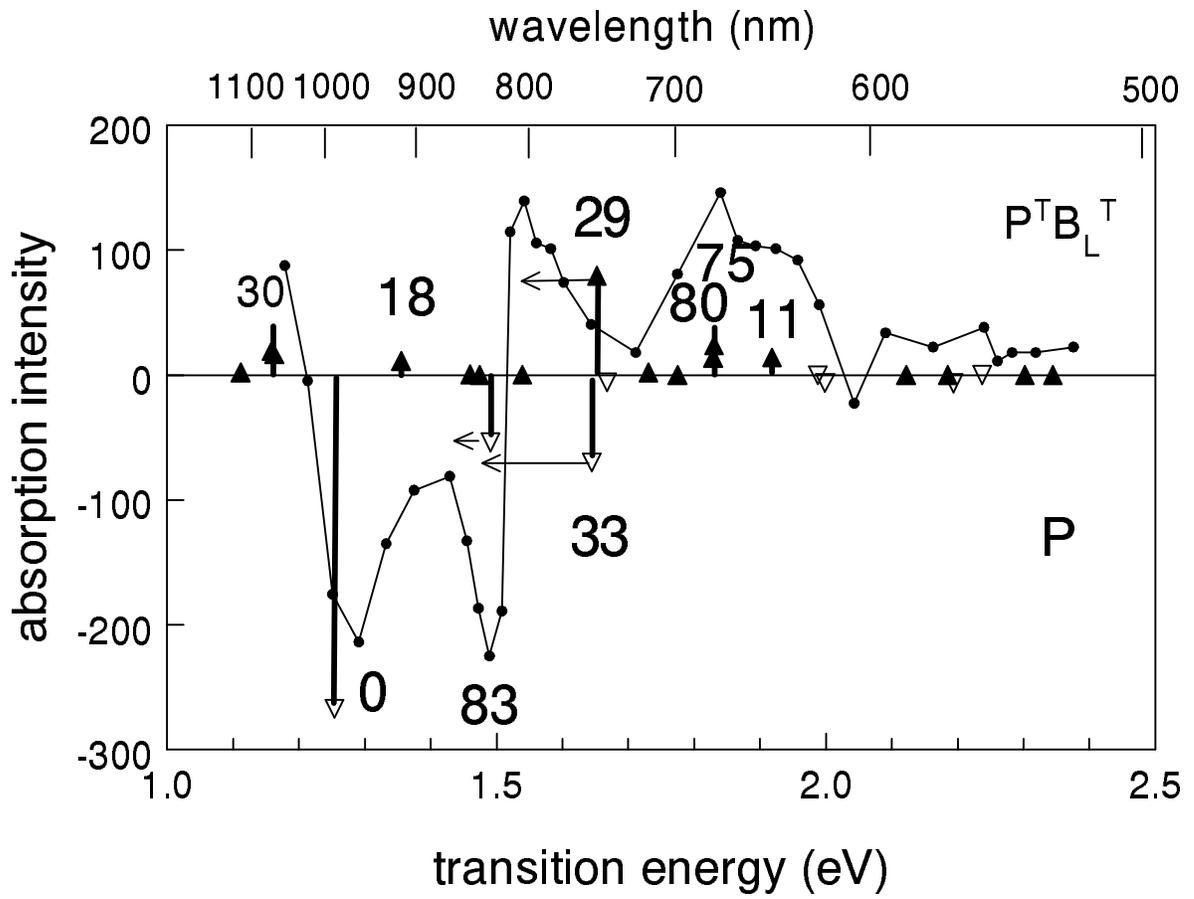
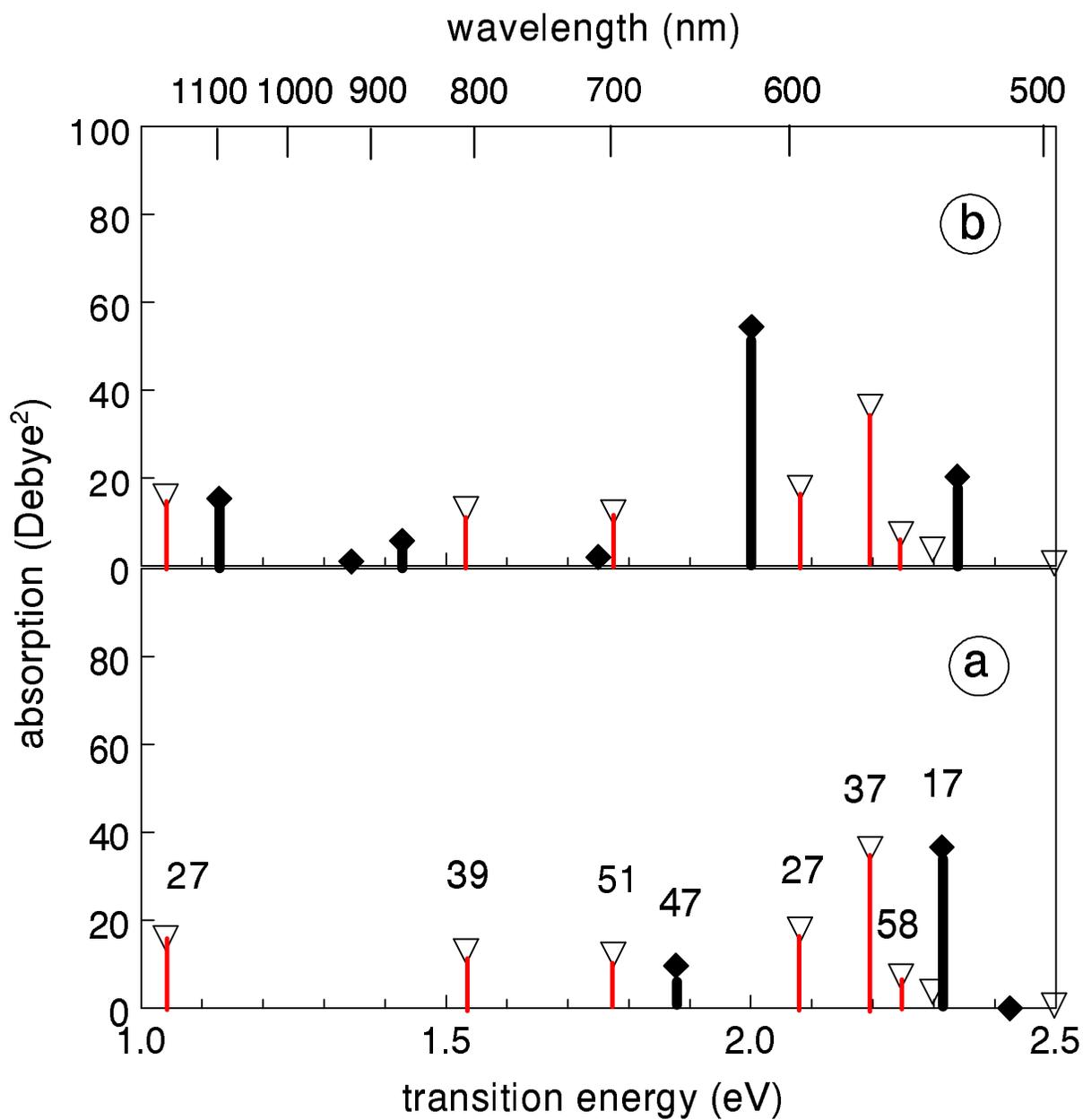
\end{document}